\documentclass[11pt]{article}
\usepackage{jheppub}
\usepackage[utf8]{inputenc}
\usepackage[T1]{fontenc}
\usepackage{epsfig,latexsym}
\usepackage{amsmath}
\usepackage[dvipsnames]{xcolor}
\usepackage{adjustbox}
\usepackage{graphicx}
\usepackage{verbatim}
\usepackage{mathrsfs}
\usepackage{amssymb}
\usepackage{multirow}
\usepackage{epsfig}
\usepackage{color,colordvi}
\usepackage{appendix}
\usepackage{slashed}
\usepackage{cancel}
\usepackage{float}
\usepackage{caption}
\usepackage{epsf}
\usepackage{amsmath}
\usepackage{physics}
\usepackage{MnSymbol}
\usepackage[normalem]{ulem}
\DeclareMathAlphabet{\mathpzc}{OT1}{pzc}{m}{it}

 \csname
@addtoreset\endcsname{equation}{section}

\def\be{\begin{equation}}
\def\ee{\end{equation}}
\def\bes{\begin{equation*}}
\def\ees{\end{equation*}}
\def\bead{\begin{aligned}}
\def\eead{\end{aligned}}
\def\bmat{\left(\begin{matrix}}
\def\emat{\end{matrix}\right)}
\def\cA{{\cal A}}
\def\cL{{\cal L}}
\def\cC{{\cal C}}

\def\cO{{\cal O}}
\def\cB{{\cal B}}
\renewcommand{\[}{\left[}
\renewcommand{\]}{\right]}
\renewcommand{\(}{\left(}
\renewcommand{\)}{\right)}

\def\ohltri{\cO_{\varphi \ell}^{(3)}}

\def\ohqtri{\cO_{\varphi q}^{(3)}}

\title{Comments on gauge anomalies at dimension-six in the Standard Model Effective Field Theory %SMEFT
}

\author[a]{Quentin~Bonnefoy,}
\author[a]{Luca~Di~Luzio,}
\author[a,b]{Christophe~Grojean,}
\author[a,b]{Ayan~Paul}
\author[a,b]{and Alejo~N.~Rossia}
\affiliation[a]{Deutsches Elektronen-Synchrotron (DESY), D-22607 Hamburg, Germany}
\affiliation[b]{Institut f{\"u}r Physik, Humboldt-Universit{\"a}t zu Berlin, D-12489 Berlin, Germany}
\emailAdd{quentin.bonnefoy@desy.de}
\emailAdd{luca.diluzio@desy.de}
\emailAdd{christophe.grojean@desy.de}
\emailAdd{ayan.paul@desy.de}
\emailAdd{alejo.rossia@desy.de}

\abstract{
We study whether higher-dimensional operators in effective field theories, in particular in the Standard Model Effective Field Theory (SMEFT), can source gauge anomalies via the modification of the interactions involved in triangle diagrams. We find no evidence of such gauge anomalies at the level of dimension-6 operators that can therefore be chosen independently to each others without spoiling the consistency of SMEFT,  at variance with recent claims.
 The underlying reason is that gauge-invariant combinations of Goldstone bosons and massive gauge fields are allowed to couple to matter currents which are not conserved. We show this in a toy model by computing the relevant triangle diagrams, as well as by working out Wess--Zumino terms in the bosonic EFT below all fermion masses.  The same approach applies directly to the Standard Model both at the renormalisable level, providing a convenient and unusual way to check that the SM is anomaly free, as well as at the non-renormalisable level in SMEFT.}

\begin{document}

\begin{flushright}
DESY 20-220\\
HU-EP-20/40
\end{flushright}
\maketitle
%\tableofcontents

%%%%%%%%%%%%%%%%%%%%%%%%%%%%%%%%%%%%%%%%%%%%%%%%%%%%%%%%%%%%%%%%%%%%
\section{Introduction}
Effective field theories (EFTs) have become a very important tool to parametrize Beyond Standard Model (BSM) physics over the last few years due to the lack of evidence of new particles close to the electroweak (EW) scale. Hence, the community has started devoting more efforts to the study of their more formal aspects. One of the best known EFTs is the Standard Model Effective Field Theory (SMEFT), which is built by adding to the Standard Model (SM) higher-dimensional gauge invariant local operators constructed with only SM fields~\cite{Grzadkowski:2010es}. Formally, its Lagrangian density is a series expansion in the parameter $1/\Lambda$, where $\Lambda$ is taken as the typical energy scale at which BSM particles lie. 

The importance of anomalies in EFTs is well-known, especially in the context of pion and axion EFTs, where the procedure of anomaly matching is used to constrain the low-energy phenomenology from the properties of the UV model and vice-versa. Additionally, any consistent gauge quantum field theory must be free of gauge anomalies (see e.g.~Ref.~\cite{Bilal:2008qx} for a review). Then, the study of anomalies in EFTs is interesting from both a formal and a phenomenological point of view. 

Gauge anomalies, a priori, count the chiral degrees of freedom and are intrinsically fixed by their quantum numbers, and not by their interactions whose relative importance is, anyway, a function of the energy of the probe.
Since the SMEFT has the same field content as the SM, one would naively expect it to be 
free of gauge anomalies.
Still, new interactions among gauge bosons and fermions arise from higher-dimensional operators and these interactions can be chiral.
So it is an interesting question to explicitly check that they are not the source of any new anomaly. 
Actually, it has been recently claimed
in Ref.~\cite{Cata:2020crs} that the SMEFT can develop gauge anomalies at 
dimension-6 and therefore, they must be cancelled order by order in the $1/\Lambda$ expansion. Such anomaly cancellation conditions would take the form of sum rules on the Wilson Coefficients (WCs) of the EFT, for instance
\begin{align}
\label{eq:condCata1}
    c_{\varphi q}^{(3)}&=  c_{\varphi l}^{(3)} \ ,\\
    \frac{c_{\varphi q}^{(1)}}{y_{q}}&=  \frac{c_{\varphi l}^{(1)}}{y_{l}} = \frac{c_{\varphi u}}{y_{u}} = \frac{c_{\varphi d}}{y_{d}} = \frac{c_{\varphi e}}{y_{e}} \ ,
    \label{eq:condCata2}
\end{align}
for the following operators of the Warsaw basis~\cite{Grzadkowski:2010es},
\be
\bead
&\cO_{\varphi \psi_R}=i\frac{c_{\varphi \psi,ij}}{\Lambda^2}\(\varphi^\dagger \overleftrightarrow D_\mu \varphi\)\overline\psi^i_R\gamma^\mu\psi^j_R \ ,\\
&\cO_{\varphi \psi_L}^{(1)}=i\frac{ c^{(1)}_{\varphi \psi,ij}}{\Lambda^2}\(\varphi^\dagger \overleftrightarrow D_\mu \varphi\)\overline\psi^i_L\gamma^\mu\psi^j_L\ , \quad \cO_{\varphi \psi_L}^{(3)}=i\frac{ c^{(3)}_{\varphi \psi,ij}}{\Lambda^2}\(\varphi^\dagger \overleftrightarrow D_\mu^a \varphi\)\overline\psi^i_L\tau^a\gamma^\mu\psi^j_L \ ,
\eead
\label{relevantSMEFTops}
\ee
(we use the conventions of Ref.~\cite{Grzadkowski:2010es}, $\psi_{L,R}$ spans over the SM chiral fields, $y_{\psi}$ is the hypercharge of the SM fields and the WCs are traced over the 3 generations generations, e.g., $c_{\varphi q}^{(3)}=\sum_i c_{\varphi q,ii}^{(3)}$).
In this note, we show that these sum rules are not necessary for the electroweak symmetry of the SM to be preserved at the quantum level. 

We start in Section~\ref{sec:TreeLevelMatch} by studying two anomaly-free UV models for the BSM sector that can be matched onto the SMEFT. The first of them violates the aforementioned sum-rules and constitutes a counterexample. On the other hand, the second model fulfils the conditions and we clarify the reason behind this. In Section~\ref{sec:CurrConsToyModel}, we study current conservation at the quantum level in a toy model with dimension-6 interactions. We identify the equivalent of Eqs.~\eqref{eq:condCata1}-\eqref{eq:condCata2} and show that such conditions are not needed for the conservation of the appropriate Noether current in triangle diagrams, either in the unbroken or broken phases. The role of Goldstone bosons (GBs) in the latter case is emphasized. We compute in a second step the Wess--Zumino (WZ) terms which appear in a low-energy bosonic EFT of GBs and gauge fields, and, consistently, we do not find any dimension-6 contribution to these terms. We apply the same logic to the SMEFT in Section~\ref{sec:SMEFTcase} and we find that the conditions in Eqs.~\eqref{eq:condCata1}-\eqref{eq:condCata2} are not necessary for the consistency of the theory. Finally, in Section~\ref{sec:conclusions} we summarize our arguments and findings. We show in Appendix~\ref{unbrokenPhaseTriangles} that the Ward-Takahashi identity holds in the unbroken phase of the toy model of Section~\ref{sec:CurrConsToyModel}, and we present in Appendix~\ref{WZappendix} some details about the derivation of WZ terms in the neutral sector of the SMEFT.

%%%%%%%%%%%%%%%%%%%%%%%%%%%%%%%%%%%%%%%%%%%%%%%%%%%%%%%%%%%%%%%%%%%%
\section{Tree-level (counter)examples}
\label{sec:TreeLevelMatch}

In this section, we review the explicit tree-level matching between two UV models and the SMEFT in order to, first, give an example of a model that violates the sum rules in Eqs.~\eqref{eq:condCata1}-\eqref{eq:condCata2} and, second, explain why certain models automatically satisfy these sum rules. We will use the conventions and results contained in Ref.~\cite{deBlas:2017xtg} that gives the complete tree-level matching of general BSM models onto dimension-6 SMEFT.

%%%%%%%%%%%%%%%%%%%%%%%%%%%%%%%%%%%%%%
\subsection{Heavy right-handed singlet Majorana fermion}
\label{sec:HeavyMajoranaFermion}

We start by considering a model that extends the SM with a heavy neutral lepton, i.e., a SM singlet Majorana fermion $N$ with a Majorana mass $M_{N}$. This is a simplified version of the see-saw type-I mechanism used to give a mass to the neutrinos and can be generalized without modifying our conclusions. The full UV Lagrangian of this model is the SM Lagrangian plus the usual kinetic and mass terms for $N$ and the following interaction terms:
\begin{equation}
	\cL_{BSM}^{\text{Int}} = - \left(\lambda_N\right)_{i} \bar{N} \tilde{\varphi}^{\dagger} \ell_{L,i},
\end{equation}
where $i=\,\,1,\,\, 2,\,\, 3 $, and we neglect higher-dimensional interaction terms which would not change our results and might obscure the discussion~\cite{deBlas:2017xtg}.

The heavy singlet fermion can be integrated out to match onto the SMEFT. This generates three classes of higher-dimensional operators with dimension not higher than 6: the Weinberg operator $\cO_{5}$ and the dimension-6 operators $\cO_{\varphi l}^{(1)}$ and $\cO_{\varphi l}^{(3)}$~\cite{deBlas:2017xtg}. More specifically, the WCs of these three operators read~\cite{deBlas:2017xtg}:
\begin{align}
	\frac{1}{\Lambda}\left(c_{5}\right)_{ij} =& \frac{\left(\lambda_{N}\right)_{j}\left(\lambda_{N}\right)_{i}}{2M_{N}},\\
	\frac{1}{\Lambda^2}\left(c_{\varphi \ell}^{(1)}\right)_{ij}=& \frac{\left(\lambda_{N}\right)_{i}^{*}\left(\lambda_{N}\right)_{j}}{4 M_{N}^{2}},\\
    \frac{1}{\Lambda^2}\left(c_{\varphi \ell}^{(3)}\right)_{ij}=& - \frac{\left(\lambda_{N}\right)_{i}^{*}\left(\lambda_{N}\right)_{j}}{4 M_{N}^{2}}.
\end{align}
On the other hand, it is clear that no operator involving quarks can be generated and hence $c_{\varphi q}^{(3)}=c_{\varphi q}^{(1)}=c_{\varphi u}=c_{\varphi d}=0$. 

Therefore, this model violates both constraints in Eqs.~\eqref{eq:condCata1}-\eqref{eq:condCata2}. At the same time, the UV renormalizable model that this EFT comes from is anomaly free because it is just the SM plus a  singlet Majorana fermion, which has no anomaly. %In other words, both the SM and BSM sectors of this model are gauge anomaly free on their own.
Furthermore, this is a vector-like extension of the SM and hence there cannot be non-decoupling BSM effects. Indeed, the additional fermion can be explicitly decoupled by sending $M_{N}\rightarrow \infty$. In conclusion, the model presented here clearly displays a violation of the sum rules in  Eqs.~\eqref{eq:condCata1}-\eqref{eq:condCata2} contrary to the claims made in Ref.~\cite{Cata:2020crs}. Other examples of violation of these sum rules can be conceived  using vector-like fermions~\cite{deBlas:2017xtg}.

%%%%%%%%%%%%%%%%%%%%%%%%%%%%%%%%%%%%%
\subsection{Heavy U(1) gauge field}
\label{sec:HeavyU1GaugeField}
Now, we will briefly review a model that has been used to validate the sum rules in Ref.~\cite{Cata:2020crs}. Let us add to the SM a massive gauge boson $\cB_{\mu}$ that comes from a $U(1)$ gauge group that commutes with the rest of the SM gauge group. We denote its mass $M_{\cB}$ and are agnostic about its ultimate origin. We suppose that this new gauge field couples to all the fermion and scalar fields of the SM in a flavour-universal manner and then its interaction terms are~\cite{deBlas:2017xtg}:
\begin{align}
	\cL_{BSM}^{\text{Int}}=& - g_{\cB}\, q_{\ell}^{\cB}\, \cB_{\mu}\, \bar{\ell}_{L}^{i}\gamma^{\mu}\ell_{L}^{i}  - g_{\cB} \, q_{q}^{\cB}\, \cB_{\mu}\, \bar{q}_{L}^{i}\gamma^{\mu}q_{L}^{i} \\ 
	& - g_{\cB}\, q_{e}^{\cB}\, \cB_{\mu}\, \bar{e}_{R}^{i} \gamma^{\mu} e_{R}^{i} - g_{\cB} \, q_{u}^{\cB}\, \cB_{\mu} \bar{u}_{R}^{i}\gamma^{\mu} u_{R}^{i} - g_{\cB} \, q_{d}^{\cB}\, \cB_{\mu}\, \bar{d}_{R}^{i}\gamma^{\mu} d_{R}^{i} \\
	& - (g_{\cB}\, q_{\varphi}^{\cB}\, \cB^{\mu}\, \varphi^{\dagger}i D_{\mu}\varphi + \text{h.c.}).
\end{align}
The conclusions of this section are independent of the flavour-universality assumption, which we make for simplicity.

This model is not automatically anomaly-free and, in particular there can be mixed anomalies between the new gauge group and the SM one. The equations that must be satisfied by the charges under $\cB$ are:
\begin{align}
    \langle \cB BB \rangle: \,\,\, & 0 = 6 q^{\cB}_{q} y_{q}^{2} + 2 q_{\ell}^{\cB} y_{\ell}^2 - q^{\cB}_{e} y_{e}^{2}-3 q^{\cB}_{u} y_{u}^{2}- 3 q^{\cB}_{d} y_{d}^{2}, \label{eq:AnomCancU1BB}\\
	\langle \cB WW \rangle: \,\,\, & 0 = 6 q^{\cB}_{q} + 2 q^{\cB}_{\ell},\label{eq:AnomCancU1WW}\\
	\langle \cB GG \rangle: \,\,\, & 0 = 2 q^{\cB}_{q} - q^{\cB}_{u} - q^{\cB}_{d},\label{eq:AnomCancU1GG}\\
	\langle \cB RR \rangle: \,\,\, & 0 = 6 q^{\cB}_{q} + 2 q_{\ell}^{\cB} - q^{\cB}_{e} - 3 q^{\cB}_{u} - 3 q^{\cB}_{d}, \label{eq:AnomCancU1RR}
\end{align}
where $R$ refers to the Ricci tensor which appears in the gravitational anomaly.

After integrating out the massive gauge boson $\cB_{\mu}$, we obtain an EFT Lagrangian with several dimension-6 operators~\cite{deBlas:2017xtg}, but not $\ohltri$ nor $\ohqtri$. Then, Eq.~\eqref{eq:condCata1} will be satisfied trivially. On the other hand, the operators involved in Eq.~\eqref{eq:condCata2} are generated with the following non-zero WCs:
\begin{align}
	\frac{1}{\Lambda^2} \( c_{\varphi l}^{(1)} \)_{ij} = & - g_{\cB}^2 \frac{q^{\cB}_{\varphi}\, q^{\cB}_{\ell}}{M_{\cB}^{2}}\delta_{ij}\\
	\frac{1}{\Lambda^2} \( c_{\varphi q}^{(1)} \)_{ij} = & - g_{\cB}^2 \frac{q^{\cB}_{\varphi}\, q^{\cB}_{q}}{M_{\cB}^{2}}\delta_{ij}\\
	\frac{1}{\Lambda^2} \( c_{\varphi e} \)_{ij} = &- g_{\cB}^2 \frac{q^{\cB}_{\varphi}\, q^{\cB}_{e}}{M_{\cB}^{2}}\delta_{ij}\\
	\frac{1}{\Lambda^2} \( c_{\varphi d} \)_{ij} = &- g_{\cB}^2 \frac{q^{\cB}_{\varphi}\, q^{\cB}_{d}}{M_{\cB}^{2}}\delta_{ij}\\
	\frac{1}{\Lambda^2} \( c_{\varphi u} \)_{ij} = &- g_{\cB}^2 \frac{q^{\cB}_{\varphi}\, q^{\cB}_{u}}{M_{\cB}^{2}}\delta_{ij},
\end{align}
and then, up to an overall common factor, there is a one-to-one correspondence between the fermion charges and the 5 non-vanishing WCs. In consequence, we can rewrite the UV anomaly conditions in Eqs.~\eqref{eq:AnomCancU1BB}-\eqref{eq:AnomCancU1RR} in terms of the WCs as follows:
\begin{align}
	0 = &\; 6 c_{\varphi q}^{(1)} y_{q}^{2} + 2 c_{\varphi l}^{(1)} y_{\ell}^2 - c_{\varphi e} y_{e}^{2}-3 c_{\varphi u} y_{u}^{2}- 3 c_{\varphi d} y_{d}^{2}, \label{eq:AnomWCU1BB}\\
	0 = &\; 6 c_{\varphi q}^{(1)} + 2 c_{\varphi l}^{(1)},\label{eq:AnomWCU1WW}\\
	0 = &\; 2 c_{\varphi q}^{(1)} - c_{\varphi u} - c_{\varphi d},\label{eq:AnomWCU1GG}\\
	0 = &\; 6 c_{\varphi q}^{(1)} + 2 c_{\varphi l}^{(1)} - c_{\varphi e} - 3 c_{\varphi u} - 3 c_{\varphi d}. \label{eq:AnomWCU1RR}
\end{align}
These are the equations found in Table 1 of Ref.~\cite{Cata:2020crs} for the special case $c_{\varphi q}^{(3)}=c_{\varphi l}^{(3)}=0$ and their only solution is Eq. (\ref{eq:condCata2}). Hence, the UV anomaly cancellation condition of the model is the reason behind the fulfilment of these conditions on the WCs.  Notice that this model belongs to the Universal Theories class because the BSM sector couples to the SM only through the scalar and fermion vector currents already present in the SM~\cite{Wells:2015uba}. It is known that those theories, when matched onto the SMEFT, always accidentally obey Eqs.~\eqref{eq:condCata1}-\eqref{eq:condCata2}~\cite{Wells:2015uba,Grojean:2018dqj}.

%%%%%%%%%%%%%%%%%%%%%%%%%%%%%%%%%%%%%%%%%%%%%%%%%%%%%%%%%%%%%%%%%%%%
\section{Current conservation at one-loop: a toy model}
\label{sec:CurrConsToyModel}

We will now explain why the sum-rules in Eqs.~\eqref{eq:condCata1}-\eqref{eq:condCata2}  are not connected to current conservation in the quantum theory, contrary to the approach advocated in Ref.~\cite{Cata:2020crs}. As we will show, the GB couplings arising from scalar fields such as the Higgs field are crucial for the analysis. 

Let us consider the following toy model of two left-handed (LH) and two right-handed~(RH) Weyl fermions, arranged in two Dirac fermions $\psi_{k=1,2}$, a complex scalar $\varphi$ and a gauge field $A_\mu$ of field strength $F_A$,
\be
\cL=-\frac{1}{4g_A^2}F_{A,\mu\nu}^2+\abs{\partial\varphi}^2-V(\abs{\varphi})+i\overline\psi_k\slashed D \psi_k +i\frac{c_{L,k}}{\Lambda^2}\(\varphi^\dagger \overleftrightarrow \partial_\mu \varphi\) \overline\psi_{k,L}\gamma^\mu\psi_{k,L}+i\frac{c_{R,k}}{\Lambda^2}\(\varphi^\dagger \overleftrightarrow \partial_\mu \varphi\) \overline\psi_{k,R}\gamma^\mu\psi_{k,R} \ ,
\label{lagToyModel}
\ee
where $D_\mu\psi_k=(\partial_\mu+iq_k^A A_\mu)\psi_k$ contains a vector-like coupling to $A$. Summation over repeated indices is implicit. Note that $\varphi$ is uncharged under the gauge symmetry $U(1)_A$ of 
$A$ and fermions are massless. We do not include bare mass terms for the fermions, although they are allowed by the gauge symmetry, since we want to study which chiral symmetries of Eq.~\eqref{lagToyModel} can be consistently gauged. For simplicity, we also postpone the inclusion of Yukawa couplings to Section~\ref{WZtoysection}. %The same remark applies to off-diagonal entries in $c_{L/R}$ which would select anomalous chiral symmetries, so we choose $c_{L/R}$ to be diagonal, $c_{L/R,kl}=\delta_{kl}c_{L/R,k}$.

%%%%%%%%%%%%%%%%%%%%%%%%%%%%%%%%%%%%%%%%%%%%%%%%%%%%%%%%%%%%%%%%%%%%
\subsection{Current (non-)conservation in triangle diagrams}\label{currentConservationSection}

Let us start by identifying the Noether currents which are classically conserved. The Noether current for $U(1)_A$ is
\be
J_\mu^A=q_k^A\overline\psi_k\gamma_\mu\psi_k  \ .
\ee
In addition, the theory has global (chiral) symmetries. Let us study the one under which $\varphi\rightarrow e^{iq_\varphi^B \epsilon_B}\varphi\, , \ \psi_k\rightarrow e^{iq_k^B\gamma_5 \epsilon_B}\psi_k$, whose associated Noether current is
\be
J_\mu^B=-iq_\varphi^B\(-\varphi^\dagger \overleftrightarrow \partial_\mu \varphi+2i\frac{c_{L,k}}{\Lambda^2}\abs{\varphi}^2 \overline\psi_{k,L}\gamma_\mu\psi_{k,L}+2i\frac{c_{R,k}}{\Lambda^2}\abs{\varphi}^2 \overline\psi_{k,R}\gamma_\mu\psi_{k,R}\)+q_k^B\overline\psi_k\gamma_\mu\gamma_5\psi_k \ .
\label{noetherB}
\ee
Using the equations of motion (eoms),
\be
\bead
&\Box \varphi= V'\(\abs{\varphi}\)+i\frac{c_{L,k}}{\Lambda^2}\partial\varphi \overline\psi_{k,L}\gamma\psi_{k,L}+i\frac{c_{L,k}}{\Lambda^2}\partial\(\varphi \overline\psi_{k,L}\gamma\psi_{k,L}\)+(L\leftrightarrow R) \ , \\
&\slashed D \psi_{k,L/R}+\frac{c_{L/R,k}}{\Lambda^2}\(\varphi^\dagger \overleftrightarrow \partial_\mu \varphi\) \gamma^\mu\psi_{k,L/R}=0 \ ,
\eead
\ee
it is easy to see that the current is conserved, $\partial^\mu J_\mu^B=0$ (actually, the parts proportional to $q_\varphi^B$ and $q_k^B$ vanish independently, since they correspond to independent symmetries of the action). Note that these remarks apply both in the broken or unbroken phase, since the eoms do not change, neither do the expressions of the currents.

Now we ask: when can we gauge the symmetry $U(1)_B$ of current $J_\mu^B$, i.e., when is it anomaly free? The usual anomaly cancellation at dimension 4 would enforce (we remind the reader that we sum over repeated indices)
\be
U(1)_A^2\times U(1)_B : \ \(q_k^A\)^2q_k^B=0 \ , \quad U(1)_B^3 : \ \(q_k^B\)^3=0 \ ,
\label{dim4AnomaliesToy}
\ee
which implies $q_1^B=-q_2^B$ and, if $q_k^B\neq 0$, $q_1^A=\pm q_2^A$. The other anomalies vanish trivially since $U(1)_A$ is vectorlike and $U(1)_B$ axial.

Had we now gauged $U(1)_B$, the Lagrangian would be
\be
\bead
\cL=-\frac{1}{4g_A^2}&F_{A,\mu\nu}^2-\frac{1}{4g_B^2}F_{B,\mu\nu}^2+\abs{D\varphi}^2-V(\abs{\varphi})+i\overline\psi_k\slashed D \psi _k\\
&+i\frac{c_{L,k}}{\Lambda^2}\(\varphi^\dagger \overleftrightarrow D_\mu \varphi\) \overline\psi_{k,L}\gamma^\mu\psi_{k,L}+i\frac{c_{R,k}}{\Lambda^2}\(\varphi^\dagger \overleftrightarrow D_\mu \varphi\) \overline\psi_{k,R}\gamma^\mu\psi_{k,R},
\label{gaugedToyModel}
\eead
\ee
where now $D_\mu\psi_k=(\partial_\mu+iq_k^A A_\mu+iq_k^B \gamma_5 B_\mu)\psi_k,\, D_\mu\varphi=(\partial_\mu+iq_\varphi^B B_\mu)\varphi$. Following the logic delineated in Ref.~\cite{Cata:2020crs}, we would find additional gauge anomalies at dimension 6. %which would imply that $c_L=c_R$, where $c_{L/R}\equiv c_{L/R,kk}$,
Indeed, we would find in the Lagrangian
\be
\cL\supset -\(q_k^B+q_\varphi^B\frac{2c_{L,k}\abs{\varphi}^2}{\Lambda^2}\)\overline \psi_{k,L}\slashed B\psi_{k,L}-\(-q_k^B+q_\varphi^B\frac{2c_{R,k}\abs{\varphi}^2}{\Lambda^2}\)\overline \psi_{k,R}\slashed B\psi_{k,R} \ ,
\label{toyModelNewCouplings}
\ee
and the logic followed in Ref.~\cite{Cata:2020crs} would demand that the expression,
\be
\tilde J_\mu^B=2q_\varphi^B\(\frac{c_{L,k}}{\Lambda^2}\abs{\varphi}^2 \overline\psi_{k,L}\gamma_\mu\psi_{k,L}+\frac{c_{R,k}}{\Lambda^2}\abs{\varphi}^2 \overline\psi_{k,R}\gamma_\mu\psi_{k,R}\)+q_k^B\overline\psi_k\gamma_\mu\gamma_5\psi_k  \ ,
\ee
which is the "current" associated to the gauge field couplings in Eq.~\eqref{toyModelNewCouplings}, is conserved in the quantum theory, at least in a broken phase where $\langle\abs{\varphi}^2\rangle=\frac{v^2}{2}\neq 0$. This is equivalent to the constraints %$c_L=c_R$ (and $q_\psi^A$ or $q_\psi^B=0$). 
\be
\bead
U(1)_A^2\times \tilde J_\mu^B : \ &\(q_k^A\)^2q_\varphi^B(c_{L,k}-c_{R,k})=0 \ ,\\
U(1)_A\times U(1)_B \times \tilde J_\mu^B : \ &q_k^Aq_k^Bq_\varphi^B(c_{L,k}+c_{R,k})=0 \ , \\
U(1)_B^2  \tilde J_\mu^B : \ &\(q_k^B\)^2q_\varphi^B(c_{L,k}-c_{R,k})=0 \ .
\eead
\label{dim6AnomaliesToy}
\ee
After imposing Eq.~\eqref{dim4AnomaliesToy}, if $q_\varphi^B,q_k^A,q_k^B\neq 0$ and $q_1^A=q_2^A$, the solution is
%$c_{L,kk}=c_{R,kk},(-1)^kc_{L,kk}+(-1)^kc_{R,kk}=0$.
$c_{L,1}=c_{R,2},c_{L,2}=c_{R,1}$. If instead, $q_\varphi^B,q_k^A\neq 0,q_k^B= 0$ or $q_1^A=-q_2^A$, the only constraint is $\sum_kc_{L,k}=\sum_kc_{R,k}$, etc.

However, $\tilde J_\mu^B$ is not even conserved in the classical theory, where it does not correspond to a Noether current. In other words, gauge invariance does not demand that the gauge invariant combination $\varphi^\dagger \overleftrightarrow D_\mu \varphi$ couples to a conserved current. The correct Noether current $J_\mu^B$ features an additional bosonic term, connected to the charge of $\varphi$, which must be included for $J_\mu^B$ to be conserved at the classical and, as we now show, quantum levels (provided Eq.~\eqref{dim4AnomaliesToy} holds).

\begin{figure}[!th]
\centering
\includegraphics[width=0.6\textwidth]{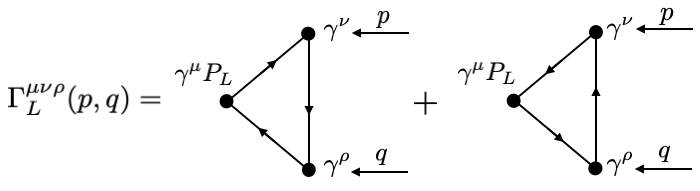}
\caption{Triangle diagrams leading to anomalies in the unbroken phase. Solid lines are fermion propagators and we indicate which combination of Dirac matrices enters each vertex.}
\label{triangleDiags}
\end{figure}

Let us analyze this claim from the point of view of the usual one-loop three-points correlation functions of symmetry currents. More precisely, let us compute
\be
\partial^\mu\langle 0 \vert J_\mu^B(x)J_\nu^A(y)J_\rho^A(z)\vert 0\rangle \ ,
\label{usualCorrelator}
\ee
the discussion being straightforwardly generalized to other combinations of currents. In the unbroken phase, the dimension-6 pieces of $J_\mu^B$ cannot be combined with the fermionic legs from the two $J_\mu^A$ to form a one-loop diagram. Indeed, that would demand closing the two Higgs legs from $J_\mu^B$ in a second loop. Consequently, at one loop the computation is the same as if the dimension-6 terms were absent and we find
\be
\text{\underline{unbroken phase}: }\partial^\mu\langle 0 \vert J_\mu^B(x)J_\nu^A(y)J_\rho^A(z)\vert 0\rangle=0%\iff q_\psi^Bq_\psi^{A,2}=0
\iff \(q_k^A\)^2q_k^B=0 \ .
\ee
Diagrammatically, defining $\Gamma_L$ as in Fig.~\ref{triangleDiags}, this arises from the fact that\footnote{To conserve the vectorlike current $J_\mu^A$, we impose $p_\nu\Gamma_L^{\mu\nu\rho}(p,q)=q_\rho\Gamma_L^{\mu\nu\rho}(p,q)=0$.} $-(p+q)_\mu \Gamma_L^{\mu\nu\rho}(p,q)=\frac{1}{4\pi^2}\epsilon^{\nu\rho\alpha\beta}p_\alpha q_\beta$ , and oppositely for $\Gamma_R$~\cite{Adler:1969gk,Bell:1969ts}.

Now let us study the broken phase. There, we parametrize $\varphi=\frac{v+h}{\sqrt 2}e^{i\frac{\theta}{v}}$, with real $v$. The Goldstone boson $\theta$ transforms under $U(1)_B$ as $\delta\theta=vq_\varphi^B\epsilon_B$, and the radial mode $h$ is inert under all symmetries. This parametrization allows us to write
\be
J_\mu^B=q_\varphi^B\(-v \partial_\mu \theta+\frac{c_{L,k}v^2}{\Lambda^2} \overline\psi_{k,L}\gamma_\mu\psi_{k,L}+\frac{c_{R,k}v^2}{\Lambda^2} \overline\psi_{k,R}\gamma_\mu\psi_{k,R}\)+q_k^B\overline\psi_k\gamma_\mu\gamma_5\psi_k +\text{ terms involving $h$} \ .
\ee
For the same reason as in the unbroken case, the terms that depend on $h$ cannot contribute to anomalies at one-loop, so we focus on the pieces that depend on the GB $\theta$ only. 

\begin{figure}[!th]
\centering
\includegraphics[width=0.7\textwidth]{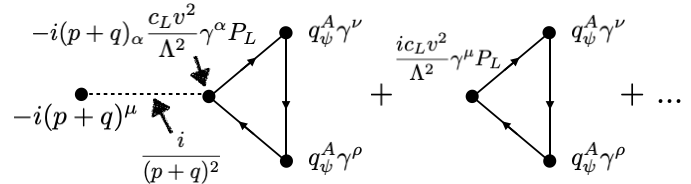}
\caption{Triangle diagrams proportional to $q_\varphi^B$
(the $+...$ refers to the same diagrams with the orientation of the fermionic arrows reversed). Solid lines are fermion propagators, the dashed one is a GB propagator and we indicate which combination of Dirac matrices and momenta enters each vertex.}
\label{triangleDiagsWithGB}
\end{figure}

Let us study the diagrams that enter the computation of $\partial^\mu\langle 0 \vert J_\mu^B(x)J_\nu^A(y)J_\rho^A(z)\vert 0\rangle$. There are diagrams proportional to $\(q_k^A\)^2q_k^B$, of the form $\(q_k^A\)^2q_k^B (\Gamma_L-\Gamma_R)$, that lead to the same contribution to the anomaly as in the unbroken phase. In addition, there are two other kinds of diagrams, proportional to $q_\varphi^B$. These are the purely fermionic diagrams sensitive to the pieces of the current proportional to $c_{L/R}$, and the ones where the current connects to a Goldstone boson propagator (see Fig.~\ref{triangleDiagsWithGB}). In the latter, there appears the three-points coupling between the GB and the fermions obtained from Eq.~\eqref{lagToyModel},
\be
\cL\supset -\frac{vc_{L,k}}{\Lambda^2} \partial_\mu \theta \overline\psi_{k,L}\gamma^\mu\psi_{k,L}-\frac{vc_{R,k}}{\Lambda^2} \partial_\mu \theta \overline\psi_{k,R}\gamma^\mu\psi_{k,R} \ .
\ee

Contracting with $-(p+q)_\mu$, it is clear that the two diagrams in Fig.~\ref{triangleDiagsWithGB} cancel, and we do not obtain any constraint on $c_{L/R}$ from those triangles. Instead, focusing only on the second diagram of Fig.~\ref{triangleDiagsWithGB} would yield the constraint $\(q_k^A\)^2q_\varphi^B(c_{L,k}-c_{R,k})=0$, as in Eq.~\eqref{dim6AnomaliesToy}, but this does not correspond to the full contribution of the Noether current. Consequently,
\be
\text{\underline{broken phase:} }\partial^\mu\langle 0 \vert J_\mu^B(x)J_\nu^A(y)J_\rho^A(z)\vert 0\rangle=0\iff \(q_k^A\)^2q_k^B=0 \ ,
\ee
and in particular we do not find any condition on $c_{L/R}$, as in the unbroken phase. As could have been expected, the IR dynamics, namely the choice of the broken or unbroken phase, does not change the conclusions regarding anomalies. Cancellations similar to that illustrated in Fig.~\ref{triangleDiagsWithGB} can also be found in the unbroken phase, as presented in Appendix~\ref{unbrokenPhaseTriangles}.

All this discussion can be summarized as follows: classically, a coupling $K^\mu(\partial_\mu\theta+vq_\varphi^BB_\mu)$ is gauge-invariant even if $K^\mu$ is not conserved, and that remains true at the quantum level.

%%%%%%%%%%%%%%%%%%%%%%%%%%%%%%%%%%%%%%%%%%%%%%%%%%%%%%%%%%%%%%%%%%%%
\subsection{Wess--Zumino terms}\label{WZtoysection}

We can rephrase the discussion in the previous section in terms of WZ terms. In the broken phase, charged chiral fermions can acquire a mass in the presence of a Yukawa term,
and one can define a bosonic EFT below their mass. There, the anomalies of the heavy fields are encoded in WZ terms~\cite{Wess:1971yu,Witten:1983tw}, for instance to ensure the overall consistency of a light, anomalous, fermionic spectrum~\cite{Preskill:1990fr}. If the anomalies of Ref.~\cite{Cata:2020crs} were genuine, they could be cancelled by adding fermions with appropriate dimension-6 couplings in the theory, and similar WZ terms should also be present to carry those new anomalies. As we are going to show, no such WZ terms are generated.

We can easily work out the WZ terms for our toy model. In particular, we gauge the anomalous symmetry associated to the current $J_\mu^B$, and we give a mass to the Weyl fermions by adding to the Lagrangian in Eq.~(\ref{gaugedToyModel}) 
the following term
\be
\delta \mathcal{L} = 
-y_{k}\varphi\overline\psi_{k,L}\psi_{k,R}+h.c. \ ,
\ee
where we chose $y_k$ real. Gauge invariance forces that $2q_k^B=q_\varphi^B$, and the Noether current in Eq.~\eqref{noetherB} respects the same anomalous Ward identities as in the case where the fermions were massless~\cite{Gross:1972pv,Bouchiat:1972iq}. 

In the broken phase, we can integrate out the fermions which obtain a mass $m_k=\frac{y_kv}{\sqrt 2}$. In the EFT below $m_k$, the anomalous WZ terms associated to triangle diagrams, irrespective of whether they arise from dimension-4 or dimension-6 vertices, are of two types: axionic terms of the form \footnote{We use the definition $\tilde{F}_j^{\mu\nu}=\frac{\epsilon^{\mu\nu\rho\sigma}}{2} F_{\rho\sigma,j}$.} $\theta F_i\tilde F_j$ and pure-gauge field terms of the form $A_i\wedge A_j\wedge F_k$, known as Generalized Chern-Simons (GCS) terms~\cite{Anastasopoulos:2006cz,Bonnefoy:2020gyh}. For our toy model, they read
\be
\cL_\text{EFT}=-\frac{C_{AA}}{16\pi^2}\frac{\theta}{v} F_A\tilde F_A-\frac{C_{AB}}{16\pi^2}\frac{\theta}{v} F_A\tilde F_B-\frac{C_{BB}}{16\pi^2}\frac{\theta}{v} F_B\tilde F_B-\frac{E_{ABA}}{8\pi^2}A_\mu B_\nu \tilde F_A^{\mu\nu}-\frac{E_{ABB}}{8\pi^2}A_\mu B_\nu \tilde F_B^{\mu\nu} \ ,
\ee
and the coefficients are
\be
C_{AA}=\(q_k^A{}\)^2 \ , \quad C_{AB}=0 \ , \quad C_{BB}=\frac{2\(q_k^B{}\)^2}{3} \ , \quad E_{ABA}=\frac{4 \(q_k^A{}\)^2 q_k^B}{3}\ , \quad E_{ABB}=0 \ .
\label{coeffsToyEFT}
\ee
They are computed from fermion triangle diagrams where the vertices are connected to $A_\mu,B_\mu,\theta$ via a dimension-4 vertex, or to the gauge-invariant contribution $iv(\partial\theta+vq_\varphi^BB)\subset \varphi^\dagger \overleftrightarrow D \varphi$ via a dimension-6 vertex (see Appendix~\ref{WZappendix} for details). A first thing to notice from Eq.~\eqref{coeffsToyEFT} is that the usual anomalies are reproduced in the EFT. Indeed,
\be
\delta\cL_\text{eff}=-\frac{\epsilon_A}{24\pi^2}2\(q_k^A\)^2q_k^BF_A\tilde F_B -\frac{\epsilon_B}{24\pi^2}\[\(q_k^A\)^2q_k^B F_A\tilde F_A+2\(q_k^B\)^3 F_B\tilde F_B\] \ ,
\ee
where $\epsilon_{A/B}$ are the parameters of the gauge transformations under which $\delta A_\mu=-\partial_\mu\epsilon_A$, $\delta B_\mu=-\partial_\mu\epsilon_B$ and $\delta\theta=vq_\varphi^B\epsilon_B$. We see there that the gauge transformations of the bosonic EFT are zero if and only if the UV anomalies in Eq.~\eqref{dim4AnomaliesToy} cancel. The second thing to notice is that there is no anomalous shift proportional to $c_{L/R}$, confirming that their values have no impact on the gauge-invariance of the theory.

%%%%%%%%%%%%%%%%%%%%%%%%%%%%%%%%%%%%%%%%%%%%%%%%%%%%%%%%%%%%%%%%%%%%
\section{The case of the SMEFT}
\label{sec:SMEFTcase}

We can now follow the same logic and show that triangle diagrams in the broken phase of the SMEFT do not constrain the WCs entering Eq.~\eqref{relevantSMEFTops}. Note that, since we restrict to triangle diagrams, we are only sensitive to the linear parts of non-abelian field strengths or to the linear parts of the gauge transformations of GBs or gauge fields. All formulae below should be read accordingly, for instance $G_{\mu\nu}^A=\partial_\mu G^A_\nu-\partial_\nu G^A_\mu$ for the gluon field strength below.

%%%%%%%%%%%%%%%%%%%%%%%%%%%%%%%%%%%%%%%%%%%%%%%%%%%%%%%%%%%%%%%%%%%%
\subsection{Gauge and Goldstone couplings}

We introduce the GBs in the Higgs field,
\be
\varphi=e^{i\frac{\pi^a}{v}\sigma^a}\bmat 0 \\ \frac{v}{\sqrt 2}\emat \ .
\ee
Under a gauge transformation $U=e^{i\frac{\sigma^a}{2}\epsilon^a+iy_\varphi\epsilon_Y}$, they transform at linear order as
\be
\delta \pi^a=\frac{v}{2}\epsilon^a-v\delta^{a3}y_\varphi\epsilon_Y \ .
\ee
We have, at linear order again,
\be
i\varphi^\dagger \overleftrightarrow D_\mu \varphi = -v^2 \(y_\varphi B_\mu -\frac{W^3_\mu}{2}\)+v\partial_\mu\pi^3\ , \quad i\varphi^\dagger \overleftrightarrow D^a_\mu \varphi = -\(\frac{v^2}{2}W^a_\mu-v^2\delta^{a3}y_\varphi B_\mu+v\partial_\mu\pi^a\) \ ,
\label{gaugeInvariantDerivativeCouplings}
\ee
and one can check that these expressions are gauge-invariant at leading order. We find the following modifications to LH gauge couplings\footnote{From those couplings, it is easy to understand the origin of the conditions in Eq.~\eqref{eq:condCata2}, equivalent to Eq.~\eqref{dim6AnomaliesToy} and presented in Ref.~\cite{Cata:2020crs}. The  gauge field-fermions couplings in Eq.~\eqref{newGaugeCouplingsSMEFT} read

\be
-\overline \psi_{L,i} \Bigg(\underbrace{\[y_\psi\delta_{ij}+v^2y_\varphi\left\{\frac{c^{(1)}_{\varphi \psi,ij}}{\Lambda^2}-2T^3\frac{c^{(3)}_{\varphi \psi,ij}}{\Lambda^2}\right\}\]}_{\equiv T_{ij}^B} \slashed B +\underbrace{T^a\[\delta_{ij}+v^2\frac{c^{(3)}_{\varphi \psi,ij}}{\Lambda^2}\]}_{\equiv T^{W,a}_{ij}}\slashed W^a\underbrace{-v^2\frac{c^{(1)}_{\varphi \psi,ij}}{2\Lambda^2}}_{\equiv T^{W_3}_{ij}}\slashed W^3\Bigg)\psi_{L,j} \ ,
\ee

where we defined new flavoured "generators" from those couplings. One can compute the anomaly polynomials associated to those new generators. Organising the result as an expansion in $\Lambda^{-2}$, they vanish at leading order since they correspond to the ones of the SM. At dimension-6, they read
\bes
\langle BBB\rangle \ : \ %3y_\varphi\frac{v^2}{\Lambda^2}\underbrace{\(6y_Q^2c^{(1)}_{\varphi Q,ii}+2y_L^2c^{(1)}_{\varphi L,ii}-3y_u^2c^{(1)}_{\varphi u,ii}-3y_d^2c^{(1)}_{\varphi d,ii}-y_e^2c^{(1)}_{\varphi e,ii}\)}_{\equiv \cA}
3y_\varphi\frac{v^2}{\Lambda^2}\(6y_Q^2c^{(1)}_{\varphi Q,ii}+2y_L^2c^{(1)}_{\varphi L,ii}-3y_u^2c_{\varphi u,ii}-3y_d^2c_{\varphi d,ii}-y_e^2c_{\varphi e,ii}\)\equiv 3y_\varphi\frac{v^2}{\Lambda^2}\cA \ ,
\ees
etc. The sum rules in Eq.~\eqref{eq:condCata2} amount to demanding that all such expressions vanish.},
\be
\bead
-\overline \psi_{L,i} \bigg(\delta_{ij}\[y_\psi \slashed B +T^a\slashed W^a\]+\[v^2 \(y_\varphi \slashed B -\frac{\slashed W^3}{2}\)-v\slashed \partial\pi^3\]&\[\frac{c^{(1)}_{\varphi \psi,ij}}{\Lambda^2}-2T^3\frac{c^{(3)}_{\varphi \psi,ij}}{\Lambda^2}\]\\
&+2T^{a=1,2}\frac{c^{(3)}_{\varphi \psi,ij}}{\Lambda^2}\[\frac{v^2}{2}\slashed W^{a}+v\slashed \partial\pi^{a}\]\bigg)\psi_{L,j} \ ,
\eead
\label{newGaugeCouplingsSMEFT}
\ee
where $T^a=\frac{\tau^a}{2}$, and for RH fields the equivalent expression is obtained by replacing $T^{a=1..3}\rightarrow 0$, $c^{(1)}_{\varphi \psi}\rightarrow c_{\varphi \psi}$ and $c^{(3)}_{\varphi \psi}\rightarrow0$.
In the Yukawa couplings, we have
\be
\bead
\cL\supset -\overline Q Y_u u \tilde \varphi-\overline Q Y_d d \varphi-\overline L Y_e e \varphi \, + h.c. =&- \frac{1}{\sqrt 2}\((v+i\pi^3)\overline u_L+i(\pi^1+i\pi^2)\overline d_L\)Y_uu_R\\
&-\frac{1}{\sqrt 2}\((v-i\pi^3)\overline d_L+i(\pi^1-i\pi^2)\overline u_L\)Y_dd_R\\
&-\frac{1}{\sqrt 2}\((v-i\pi^3)\overline e_L+i(\pi^1-i\pi^2)\overline \nu_L\)Y_ee_R+h.c.
\eead
\ee

%%%%%%%%%%%%%%%%%%%%%%%%%%%%%%%%%%%%%%%%%%%%%%%%%%%%%%%%%%%%%%%%%%%%
\subsection{Wess--Zumino terms in the SMEFT}

We now study the one-loop WZ terms generated by matching between the SMEFT and the bosonic theory below the mass of all fermions. Since anomalies are independent of fermion masses, we work in the theory with the field content and operators of the SMEFT, but with gauge and Yukawa couplings such that it makes sense to integrate out all fermions and remain with a bosonic EFT. Indeed, the quantum consistency of the theory does not depend on the precise numerical values of the Yukawas and the gauge couplings, so we are free to study anomalies in this ``deformed'' theory. In order to account for the neutrinos in the same way, we add RH neutrinos $\nu_R$ to give neutrinos a mass:
\be
\cL\supset -\overline L Y_\nu \nu_R \tilde \varphi+h.c.=- \frac{1}{\sqrt 2}\((v+i\pi^3)\overline \nu_L+i(\pi^1+i\pi^2)\overline e_L\)Y_\nu \nu_R+h.c. \ ,
\ee
but we do not include any dimension-6 coupling involving those RH neutrinos. 

Let us first remind the reader what such a procedure yields for the SM at dimension 4. In the effective theory obtained from the SM below the mass of all the fermions, focusing henceforth on neutral bosons and GBs (the charged ones can be treated identically), one finds the following couplings (see Appendix~\ref{WZappendix}),
\be
\bead
\cL_\text{EFT}\supset &-\frac{1}{16\pi^2}\frac{\pi^3}{v}B\tilde B\(3\[y_u^2+y_Q y_u-y_d^2-y_Qy_d\]+y_\nu^2+y_Ly_\nu-y_e^2-y_e y_L\)\\
&-\frac{1}{16\pi^2}\frac{\pi^3}{v}B\tilde W^3\(\frac{3(y_d+4y_Q+y_u)}{2} + \frac{y_e+4y_L+y_\nu}{2} \)\\
&-\frac{1}{8\pi^2}B_\mu W^3_\nu \tilde B^{\mu\nu}\frac{( y_\nu-y_e) (y_e + y_L + y_\nu)+3 (y_u - y_d) (y_d + y_Q + y_u)}{2}\\
&-\frac{1}{8\pi^2}B_\mu W^3_\nu \tilde W^{3,\mu\nu}\frac{3 (y_u+y_d) + y_e + y_\nu}{4} \ ,
\eead
\ee
where as we said we restrict to the linear pieces in the non-abelian field strengths, and we use the same letter to designate both the SM gauge fields and their field strengths. The variation of the action reads (using relations such as $y_u=y_Q+y_\varphi$, and similarly for all fermions),
\be
\bead
\delta \cL_\text{EFT}&=-\frac{\epsilon_Y}{16\pi^2}\[\(6y_Q^3+ 2 y_L^3- 3y_u^3-3y_d^3 - y_e^3  - y_\nu^3\)B\tilde B +\frac{3y_Q+y_L}{2}W^3\tilde W^3\]-\frac{\epsilon_3}{16\pi^2}\(3y_Q+y_L\)B\tilde W^3.
\eead
\ee
Demanding that those variations vanish, we recover the known $U(1)_Y^3$ and $U(1)_Y\times SU(2)_L^2$ anomaly cancellation conditions.

Now we can study the WZ terms which arise from inserting one dimension-six vertex in the triangle diagrams. They are all found to vanish (see Appendix~\ref{WZappendix} for details). As for the toy model in Section~\ref{sec:CurrConsToyModel}, the SM fermions do not generate any breakdown of gauge invariance in the EFT from triangle diagrams, and in particular we do not find any sum-rule to enforce at dimension-6.

%%%%%%%%%%%%%%%%%%%%%%%%%%%%%%%%%%%%%%%%%%%%%%%%%%%%%%%
\section{Conclusions}
\label{sec:conclusions}

Chiral triangle loops are known to potentially break gauge symmetries at the quantum level. These possible  breakdown effects are RG-independent, therefore it is anticipated that, if a UV model extending the SM at high energies is anomaly free, its EFT description at lower energies when the heavy degrees of freedom are integrated out and traded off for higher dimensional interactions among the SM fields should exhibit a full $SU(3)\times SU(2) \times U(1)$ unbroken gauge symmetry.

In this note, we showed that the constraints on the WCs claimed in Ref.~\cite{Cata:2020crs} to be necessary to ensure gauge invariance at the quantum level do not hold. 
%\ANR{I think this is interesting, but at the same time we're not sure whether this happens always as a consequence of UV gauge anomaly cancellation in those models.}
%
Instead, we explained how to consistently compute the gauge anomalies in EFTs and stressed the importance of including the GBs in the computation. With a very simple toy model, we showed that their inclusion is a natural consequence of the proper identification of the classical current that may be anomalous. In other words, ignoring the GBs in the gauge anomaly computation equates to asking that a classically non-conserved current be conserved at the quantum level\footnote{Let us make a remark about using the unitary gauge: fixing the GBs to zero in a Noether current and in all eoms breaks the current conservation at the classical level already. This is easily seen in a model where a fermionic mass is obtained after spontaneous symmetry breaking. Consequently, it is not expected that the current is conserved in the unitary gauge at the quantum level.
%More generally, we should not ask for the conservation of a gauge current after gauge-fixing, because gauge-fixing is an inherently gauge-breaking step. Instead, what remains after gauge-fixing is a BRST symmetry.
}. We also rephrased our discussion in terms of WZ terms, and showed that no EFT terms made out of GBs and gauge fields match the would-be dimension-6 anomalies, both in our toy model and in the (neutral sector of the) SMEFT. 

\paragraph{Note added.} 
Slightly after our work,  
Ref.~\cite{Feruglio:2020kfq} appeared 
on the arXiv reaching conclusions 
similar to ours by working in the unbroken phase 
and allowing for the most general bosonic 
background.

\acknowledgments
We thank Fady Bishara, Emanuele Gendy, Di Liu, and Philipp Englert for useful discussions. 
This work is supported by the Deutsche Forschungsgemeinschaft under Germany's Excellence Strategy  EXC 2121 ``Quantum Universe'' - 390833306.
The work of C.G.~and A.R.~was also supported by the International Helmholtz-Weizmann Research School for Multimessenger Astronomy, largely funded through the Initiative and Networking Fund of the Helmholtz Association.
The work of L.D.L.~is supported by the Marie Sk\l{}odowska-Curie 
Individual Fellowship grant AXIONRUSH (GA 840791).
%%%%%%%%%%%%%%%%%%%%%%%%%%%%%%%%%%%%%%%%%%%%%%%%%%%%%%%
\begin{appendices}

%%%%%%%%%%%%%%%%%%%%%%%%%%%%%%%%%%%%%%%%%%%%%%%%%%%%%%%%%%%%%%%%%%%%
\section{Ward-Takahashi identity in the unbroken phase}\label{unbrokenPhaseTriangles}

In order to confirm that there are no anomalies coming from the kind of dimension-6 terms studied in this note, it is useful to display cancellations similar to those in Fig.~\ref{triangleDiagsWithGB}, but in the unbroken phase of our toy model of Section~\ref{sec:CurrConsToyModel}. This means that we need to study a correlator which is sensitive (at one-loop) to an insertion of the dimension-6 part of the current in Eq.~\eqref{noetherB} (as we said in Section~\ref{sec:CurrConsToyModel}, this is not the case for the usual correlator in Eq.~\eqref{usualCorrelator}). The simplest such correlator is
\be
\langle 0 \vert J_\mu^B(x) J_\nu^A(y) J_\rho^A(z) \varphi(x_1)\varphi^\dagger(x_2)\vert 0\rangle \ ,
\label{correlatorScalars}
\ee
and the Ward-Takahashi (WT) identity which follows from the classical symmetry in the absence of anomalies is
\be
\bead
\partial^\mu\langle 0 \vert J_\mu^B(x) J_\nu^A(y)& J_\rho^A(z) \varphi(x_1)\varphi^\dagger(x_2)\vert 0\rangle \\
&+q_\varphi^B\[\delta^{(4)}(x-x_1)-\delta^{(4)}(x-x_2)\]\langle 0 \vert J_\nu^A(y) J_\rho^A(z) \varphi(x_1)\varphi^\dagger(x_2)\vert 0\rangle=0 \ ,
\label{WTidentity}
\eead
\ee
where contact terms are present since $\varphi^{(\dagger)}$ is charged under $U(1)_B$, unlike $J_{\nu(\rho)}^A$.

\begin{figure}[!th]
\centering
\includegraphics[width=0.8\textwidth]{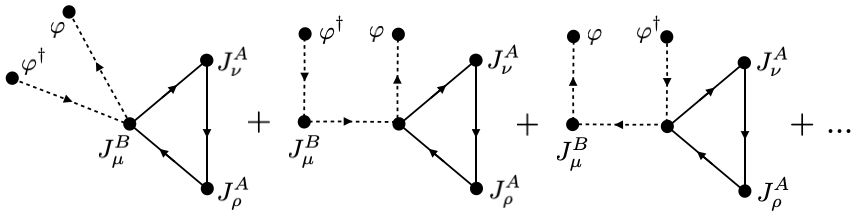}\\
\caption{Triangle diagrams  proportional to $q_\varphi^B$ entering the correlator in Eq.~\eqref{correlatorScalars} in the unbroken phase
(the $+...$ refers to the same diagrams with the orientation of the fermionic arrows reversed). Solid lines are fermion propagators and dashed lines are scalar propagators.}
\label{unbrokenPhaseDiags1}
\end{figure}

Let us compute the correlator in Eq.~\eqref{correlatorScalars} at one-loop, focusing on the diagrams proportional to $q_\varphi^B$. They are displayed in Fig.~\ref{unbrokenPhaseDiags1}.

\begin{figure}[!th]
\centering
\includegraphics[width=0.4\textwidth]{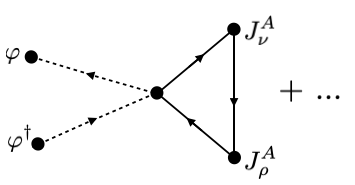}\\
\caption{Triangle diagrams entering the second correlator in Eq.~\eqref{WTidentity} in the unbroken phase
(the $+...$ refers to the same diagrams with the orientation of the fermionic arrows reversed). Solid lines are fermion propagators and dashed lines are scalar propagators.}
\label{unbrokenPhaseDiags2}
\end{figure}

\noindent We obtain
\be
\bead
\langle 0 \vert J_\mu^B(x)& J_\nu^A(y) J_\rho^A(z) \varphi(x_1)\varphi^\dagger(x_2)\vert 0\rangle\\
&=iq_\varphi^B\frac{c_L}{\Lambda^2}\int \frac{d^4p \, d^4q}{(2\pi)^{8}}\frac{e^{i\[p \cdot (x-x_1) +q \cdot (x-x_2)\]}}{(p^2-m_\varphi^2)(q^2-m_\varphi^2)}(D_{L,1}+D_{L,2}+D_{L,3})^{\mu\nu\rho}+(L\leftrightarrow R) \ ,
\eead
\ee
where the $D_{L,i}$s correspond to the individual diagrams in Fig.~\ref{unbrokenPhaseDiags1}. They respectively read
\be
\bead
&D_{L,1}^{\mu\nu\rho} \equiv-2\Gamma_L^{\mu\nu\rho}(x,y,z)\ ,\\
&D_{L,2}^{\mu\nu\rho} \equiv \int \frac{d^4 \tilde x \, d^4l}{(2\pi)^4} e^{i(l+p)\cdot (\tilde x-x)}\frac{l^\mu+q^\mu}{l^2-m_\varphi^2}(l-p)_\alpha\Gamma_L^{\alpha\nu\rho}(\tilde x,y,z)\ , \\
&D_{L,3}^{\mu\nu\rho}\equiv D_{L,2}^{\mu\nu\rho}\big\vert_{p\leftrightarrow q, x_1 \leftrightarrow x_2} \ ,
\eead
\ee
where 
\be
\Gamma_L^{\mu\nu\rho}(x,y,z)\equiv\langle 0 \vert \overline \psi_L \gamma^\mu \psi_L (x) J_\nu^A(y) J_\rho^A(z)\vert 0\rangle
\ee
refers to the fermion loop, about which we just need to know its dependence on the spacetime points and momenta. The other correlator which enters the WT identity in Eq.~\eqref{WTidentity} receives contributions from the diagrams in Fig.~\ref{unbrokenPhaseDiags2}, and reads
\be
\bead
\langle 0 \vert J_\nu^A(y)& J_\rho^A(z) \varphi(x_1)\varphi^\dagger(x_2)\vert 0\rangle\\
&=\frac{c_L}{\Lambda^2}\int \frac{d^4p \, d^4 q \, d^4\tilde x}{(2\pi)^8}\frac{e^{i\[(p \cdot (\tilde x-x_1)+q\cdot(\tilde x- x_2)\]}}{(p^2-m_\varphi^2)(q^2-m_\varphi^2)}(p-q)_\alpha \Gamma_L^{\alpha\nu\rho}(\tilde x,y,z) +(L\leftrightarrow R) \ .
\eead
\ee

From these expressions, it is straightforward to check that the WT identity is verified, irrespective of the value (and the regularization) of the fermion loop\footnote{As can be inferred from our computation, the WT indentity would hold for any function $\Gamma_{L(R)}$ - or equivalently, for any gauge-invariant current coupled to $\varphi^\dagger \overleftrightarrow \partial \varphi$ in the lagrangian.} $\Gamma_{L(R)}^{\mu\nu\rho}$.

%%%%%%%%%%%%%%%%%%%%%%%%%%%%%%%%%%%%%%%%%%%%%%%%%%%%%%%%%%%%%%%%%%%%
\section{Derivation of the Wess--Zumino terms}\label{WZappendix}

\begin{figure}[!th]
\centering
\includegraphics[width=0.8\textwidth]{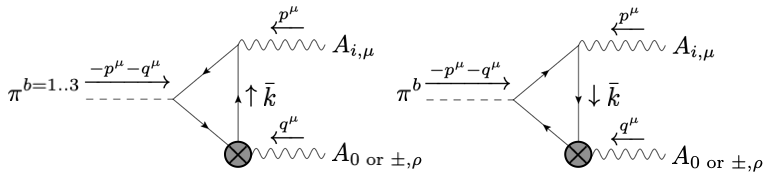}\\
\includegraphics[width=0.8\textwidth]{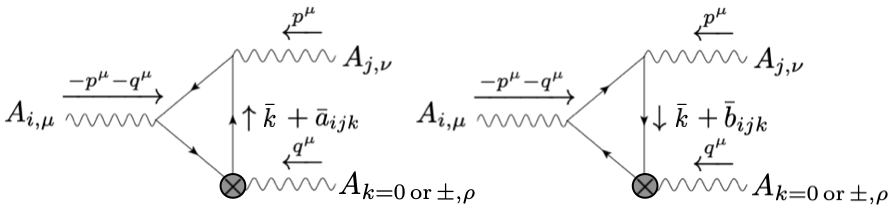}
\caption{Triangle diagrams leading to anomalous EFT terms. A black circled cross indicates a dimension-six EFT coupling. The vectors $\bar a$ and $\bar b$ correspond to shift ambiguities of the loop momentum $\bar k$ inherent to the second kind of diagrams.}
\label{axionDiags}
\end{figure}

To derive the WZ terms in the bosonic EFTs encoutered previously, we use the formulae in Appendix A of Ref.~\cite{Bonnefoy:2020gyh}. They can be immediately applied to derive the contribution of the dimension-4 couplings\footnote{Diagrams which lead to GCS terms are linearly divergent: we fix the ambiguity in momentum shifts by treating all gauge fields symmetrically when all vertices connect to dimension-4 fermion currents, and by demanding that the dimension-4 currents are conserved when one of the vertices connects to a dimension-six term. Such choices are related to counterterms in the EFT, so they do not matter when discussing relevant anomalies~\cite{Bilal:2008qx}.}, and the dimension-6 contributions can be obtained by applying the same formulae with an additional ``gauge field'', corresponding to the suitable combination of Goldstone bosons and actual gauge fields. In the toy model of Section~\ref{sec:TreeLevelMatch}, this is
\be
\hat B_\mu\equiv v(\partial_\mu \theta+vq_\varphi B_\mu) \ ,
\ee
while in SMEFT we define 
\be
A_{0,\mu}\equiv v^2 \(y_\varphi B_\mu -\frac{W^{3}_\mu}{2}\)-v\partial_{\mu}\pi^3 \ , \quad A_{\pm,\mu}\equiv \frac{v^2}{2}W^{\pm}_{\mu}+v\partial_{\mu}\pi^{\pm} \ ,
\ee
and their field strengths
\be
F_{0,\mu\nu}= v^2 \(y_\varphi B_{\mu\nu} -\frac{W^{3}_{\mu\nu}}{2}\) \ , \quad F_{\pm,\mu\nu}= \frac{v^2}{2}W^{\pm}_{\mu\nu}
\ee
(we remind the reader that we work at linear order in non-abelian gauge fields). We only present formulae related to the neutral sector of the SMEFT below, the equivalent ones for our toy model yield the results in Eq.~\eqref{coeffsToyEFT}. Anomalous EFT couplings at dimension-6 arise from the diagrams in Fig.~\ref{axionDiags}.

Let us start with the diagrams in the upper part of Fig.~\ref{axionDiags}, those with one dimension-4 insertion of a Goldstone coupling. They lead to the following EFT couplings,
\be
\cL_\text{EFT}\supset -\frac{1}{16\pi^2}\cC_{F_0B}\frac{\pi^3}{v}F_{0}\tilde B-\frac{1}{16\pi^2}\cC_{F_0W^3}\frac{\pi^3}{v}F_{0}\tilde W^3
\label{eq:EFT1}
\ee
with
\be
\bead
\cC_{F_0B}=-\frac{1}{3\Lambda^2}\Big[&3 \(c_{\varphi d}^{(1)} (2 y_d+y_Q)-c_{\varphi u}^{(1)} (y_Q+2 y_u)+c_{\varphi Q}^{(1)} (y_d-y_u)+c_{\varphi Q}^{(3)} (y_d+4 y_Q+y_u)\)\\&\qquad\qquad+c_{\varphi e}^{(1)} (2 y_e+y_L)+c_{\varphi L}^{(1)} (y_e-y_\nu)+c_{\varphi L}^{(3)} (y_e+4 y_L+y_\nu)\Big]\\
\cC_{F_0W^3}=\frac{1}{6\Lambda^2} \Big[&3c_{\varphi d}^{(1)}+3c_{\varphi u}^{(1)}+12c_{\varphi Q}^{(1)}+c_{\varphi e}^{(1)}+4c_{\varphi L}^{(1)}\Big]\\
\eead
\label{axiondim4}
\ee
To this, we should add the triangle diagrams with two insertions of dimension-4 gauge currents, those in the lower part of Fig.~\ref{axionDiags}. This leads to the following EFT terms,
\be
\bead
\cL_\text{EFT}\supset & -\frac{E_{A_0BB}}{8\pi^2\Lambda^2}A_{0,\mu} B_\nu \tilde B^{\mu\nu}-\frac{E_{A_0W^3B}}{8\pi^2\Lambda^2}A_{0,\mu} W^3_\nu \tilde B^{\mu\nu}-\frac{E_{A_0BW^3}}{8\pi^2\Lambda^2}A_{0,\mu} B_\nu \tilde W^{3,\mu\nu}\\
&-\frac{E_{BW^3F_0}}{8\pi^2\Lambda^2}B_{\mu} W^3_\nu \tilde F_0^{\mu\nu}-\frac{E_{A_0W^3W^3}}{8\pi^2\Lambda^2}A_{0,\mu} W^3_\nu \tilde W^{3,\mu\nu}
\label{eq:EFT2}
\eead
\ee
with
\be
\bead
E_{A_0BB}=&c_{\varphi u}^{(1)} (y_Q-y_u) (y_Q+2 y_u)+c_{\varphi d}^{(1)} (y_Q-y_d) (y_Q+2 y_d)- c_{\varphi Q}^{(1)} \left(y_d^2+y_d y_Q-4 y_Q^2+y_Q y_u+y_u^2\right)\\
&-c_{\varphi Q}^{(3)} (y_d-y_u) (y_d+y_Q+y_u)+\frac{1}{3} c_{\varphi e}^{(1)} (y_L-y_e) (y_L+2 y_e)\\
&-\frac{1}{3} c_{\varphi L}^{(1)} \left(y_e^2+y_e y_L-4 y_L^2+y_Ly_\nu+y_\nu^2\right)-\frac{1}{3} c_{\varphi L}^{(3)} (y_e-y_\nu)(y_e+y_L+y_\nu)\\
E_{A_0W^3B}=&\frac{1}{2}c_{\varphi u}^{(1)} (y_Q-y_u)+\frac{1}{2}c_{\varphi d}^{(1)} (y_d-y_Q)+ c_{\varphi Q}^{(1)} \left(y_d- y_u\right)+c_{\varphi Q}^{(3)} (y_d-2y_Q+y_u)\\
&+\frac{1}{6} c_{\varphi e}^{(1)} (y_e-y_L)+\frac{1}{3} c_{\varphi L}^{(1)} \left(y_e-y_\nu\right)+\frac{1}{3} c_{\varphi L}^{(3)} (y_e-2y_L+y_\nu)\\
E_{A_0BW^3}=&\frac{1}{2}c_{\varphi u}^{(1)} (y_Q+2y_u)-\frac{1}{2}c_{\varphi d}^{(1)} (y_Q+2y_d)-\frac{1}{2} c_{\varphi Q}^{(1)} \left(y_d- y_u\right)-\frac{1}{2}c_{\varphi Q}^{(3)} (y_d+4y_Q+y_u)\\
&-\frac{1}{6} c_{\varphi e}^{(1)} (y_L+2y_e)-\frac{1}{6} c_{\varphi L}^{(1)} \left(y_e-y_\nu\right)-\frac{1}{6} c_{\varphi L}^{(3)} (y_e+4y_L+y_\nu)\\
E_{BW^3F_0}=&\frac{3}{2}c_{\varphi u}^{(1)} y_u-\frac{3}{2}c_{\varphi d}^{(1)} y_d-\frac{3}{2} c_{\varphi Q}^{(1)} \left(y_d- y_u\right)-\frac{3}{2}c_{\varphi Q}^{(3)} (y_d+y_u)\\
&-\frac{1}{2} c_{\varphi e}^{(1)} y_e-\frac{1}{2} c_{\varphi L}^{(1)} \left(y_e-y_\nu\right)-\frac{1}{2} c_{\varphi L}^{(3)} (y_e+y_\nu)\\
E_{A_0W^3W^3}=&\frac{1}{12}\(3c_{\varphi u}^{(1)} +3c_{\varphi d}^{(1)} +12c_{\varphi Q}^{(1)} +c_{\varphi e}^{(1)} +4c_{\varphi L}^{(1)}\)
\eead
\label{gcsdim4}
\ee
We can rearrange those expressions into Goldstone couplings and GCS terms by separating the terms in $A_{0,\mu}$, and using integration by parts. After this rearrangement, and using only relations required by the classical gauge invariance of the theory such as $y_u=y_Q+y_\varphi$, all the EFT terms in Eqs.~(\ref{eq:EFT1}) and~(\ref{eq:EFT2}) cancel each others up to some total derivative terms. 

This result could have been anticipated: the only objects which can be formed using axionic and GCS terms, and which do not spoil the gauge-invariance of the bosonic EFT, have the following schematic form \cite{Anastasopoulos:2006cz,Bonnefoy:2020gyh}:
\be
(\partial_\mu\theta_i-A_{i,\mu})(\partial_\nu\theta_j-A_{j,\mu})\tilde F_k^{\mu\nu} \ ,
\label{invariantGCS}
\ee
where the gauge transformations of $\theta_i$ are such that $\partial_\mu\theta_i-A_{i,\mu}$ is gauge-invariant (in other words, $\theta_i$ corresponds to the longitudinal component of $A_i$). For the expression in Eq.~\eqref{invariantGCS} not to vanish, one needs at least two different massive gauge fields due to the antisymmetric structure of $\tilde F^{\mu\nu}_k$. However, there is only one massive gauge field, the $Z$ boson, in the neutral sector of the SMEFT. Therefore, any non-vanishing WZ term in the bosonic EFT must break gauge invariance, but we do not find any such term at dimension-6.

%%%%%%%%%%%%%%%%%%%%%%%%%%%%%%%%%%%%%%%%%%%%%%%%%%%%%%%
\end{appendices}

\bibliographystyle{JHEP}
\bibliography{biblio.bib}

\end{document}